\documentstyle[epsfig]{aipproc}

\begin{document}
\title{Precise Physics of Simple Atoms}

\author{Savely G. Karshenboim}
\address{D.I. Mendeleev Institute for Metrology, 198005 St. Petersburg, Russia\\
Max--Planck--Institut f\"ur Quantenoptik, 85748 Garching, Germany\thanks{Summer address}
}

\maketitle

\begin{abstract}
We give a review of experimental and theoretical results on the precision study of hydrogen--like atoms with low value of the nuclear charge $Z$.
\end{abstract}


The simplicity of ``simple'' atoms has been for a while a challenge to
precision theory and experiment. Are the hydrogen--like atomic
systems simple enough to be calculated with an accuracy, appropriate to
compete with the best experimental results? That is a question, that
theorists have tried to answer. The simplest atoms are different
two--body bound systems with a low value of the nuclear charge: $Z=1$
(hydrogen, deuterium, muonium and positronium) and $Z=2$ (ions of helium--3 and helium--4) etc. We do not try to review theoretical calculations (if necessary details can be found in Ref. \cite{Eides}), but present state of art in physics of simple atoms and discuss in detail the theoretical and experimental status of studying such atoms. 

\section*{Low-energy tests of QED}

The precise physics of simple atom is the most interesting part of the so-called low-energy tests of Quantum Electrodynamics ({\em QED}). Low energy tests of QED offer a number of different options:
\begin{itemize}
\item A study with free particles provides the possibility of testing the QED Lagrangian for free particles. The most accurate data arise from anomalous magnetic moments of the electron ({\em Kinoshita}$^\dag$\footnote{References marked with $\dag$ correspond to presentations at a satellite meeting to ICAP named {\em Hydrogen atom, 2: Precision physics of simple atomic system} and its Proceedings will be published by Springer in 2001.}) and the muon \cite{Hughes}.
\item
However, one knows that the bound problem makes all calculations more complicated. Bound state QED is not a well--established theory. It involves different effective approaches to solve a two--body problem. These approaches can be essentially checked with low $Z$ atomic systems like for e. g. hydrogen and deuterium ({\em H\"ansch}$^\dag$), neutral helium ({\em Drake}$^\dag$) and helium ions, muonium ({\em Jungmann}$^\dag$), positronium ({\em Conti}$^\dag$), muonic hydrogen and helium etc. We consider here most of the low--$Z$ atoms.
\item
Study of high--$Z$ ions ({\em Myers}$^\dag$, {\em St\"olker}$^\dag$) cannot further the test of the bound state QED because of large contributions due to the nuclear structure. Rather such an investigation is useful for trying different nuclear models. However, in some particular cases, atomic systems with a not too high $Z$ can give some important information on higher order terms of the QED $Z \alpha$ expansion. 
\item
There are some other two--body atoms under investigation. They contain a hadron as an orbiting particle. Different antiprotonic ({\em Yamazaki}$^\dag$) and pionic ({\em Nemenov}$^\dag$) atoms provide a unique opportunity to study particle property with spectroscopic means with a high precision. In some sense it is not possible to have low precision: if a signal is detected the accuracy is granted.
\end{itemize}
The precision study of the simple atoms is not only limited by experiments with simple atoms. The theory is not able to predict anything to be comparable to the experimental data. What theory can do is to express a measurable quantity in terms of fundamental constants and particle (or nuclear) properties. 

First of all we need to determine somehow the Rydberg constant ($R_\infty$), the fine structure constant ($\alpha$) and the electron mass in some appropriate units (e. g. in atomic units or in terms of the proton mass ($m_e/m_p$)). Uncertainties arising from these constants are sometimes compatible with other items of the uncertainty budget or they are even sometimes the most important source of inaccuracy. One should remember that the electron is the most fundamental particle for physics, chemistry, and metrology and the constants associated with its properties go through any atomic spectroscopic effects and any quantum electromagnetic effects. Due to that a number of different studies, which are very far from the spectroscopy of simple atoms (like e. g. Watt balance experiment (see {\em Mohr}$^\dag$ for detail)), are really strongly connected with the precision physics of simple atoms.

However, a knowledge of the universal fundamental constants is not enough for precision theoretical predictions and we need to learn also some more specific constants like for e. g. the muon mass or the proton electric charge radius. The former is important for the muonium hyperfine structure, while the latter is for calculating the hydrogen Lamb shift.

\section*{Spectra of simple atoms}

\begin{figure}[ht] 
\centerline{\epsfig{file=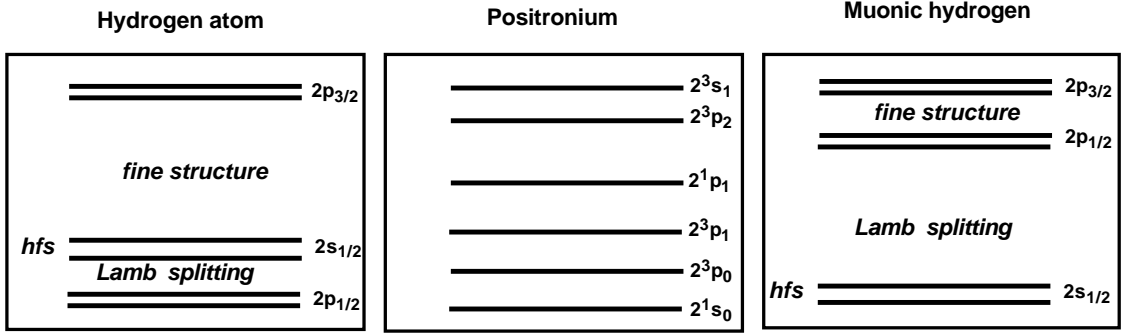,width=\textwidth}}
\vspace{10pt}
\caption{Scheme of the lowest excited levels ($n=2$) in different simple atoms}
\label{f:sgk1}
\end{figure}
Let us discuss the spectrum of simple atoms in more detail. The gross structure of atomic levels in a hydrogen--like atom comes from the Sch\"odinger equation with the Coulomb potential and the result is well--known\footnote{We use the relativistic units in which $\hbar=c=1$.} $E_n=-(Z\alpha)^2m_e$. There are a number of different corrections: the relativistic ones (one can find them from the Dirac equation), the hyperfine structure (due to the nuclear magnetic moment) and the QED ones. A structure of levels with the same value of the principal quantum number $n$ is a signature of any atomic system. In Fig. 1 we present three different spectra of the structure at $n=2$. The first one is realized in ``normal'' (electronic) hydrogen--like atoms (hydrogen, deuterium, helium ions etc). The muonium spectrum is the same. The largest splitting, of order $(Z\alpha)^4m_e$, is the fine structure (i. e. the splitting between levels with a different value of the electron angular momentum $j$), the Lamb shift arising from the electron self--energy effects is of order $\alpha(Z\alpha)^4m_e\ln\big(1/(Z\alpha)\big)$ and it splits the levels with the same $j$ and different values of the electron orbital momentum $l$. Some nuclei are spinless (like e. g. in $^4$He), while others have a non--zero spin (in hydrogen, deuterium, muonium, helium--3). In the latter case, the nuclear spin splits levels with the same electronic quantum number. The splitting are of order $(Z\alpha)^4m_e^2/M$ or $\alpha(Z\alpha)^3m_e^2/m_p$, where $M$ is the nuclear mass, and the structure depends on the value of the nuclear spin. The scheme in Fig. 1 is for nuclear spin 1/2 (hydrogen). 

The structure of levels in positronium and muonic atoms is different because some other effects enter into consideration. For positronium, an important feature is a real (into two and three photons) and virtual (into one photon) annihilation. The former is responsible for the decay of the s-states, while the latter shifts triplet levels (and 2$^3$s$_1$ in particular). The shift is of the order of $\alpha^4m_e$. Contributions of the same order arise from relativistic effects and hyperfine interactions. As a result the positronium level structure at $n=2$ has no hierarchy (Fig. 1). 

Another situation is that for the muonic atoms. A difference comes from a contribution due to the vacuum polarization effect (the Uehling potential). Effects of electronic vacuum polarization shift all levels to the order of $\alpha(Z\alpha)^2m_\mu$. This shift is a nonrelativistic one and it splits 2s and 2p levels. The fine and hyperfine structures are of the same form as for the normal atoms (i. e. $(Z \alpha)^4m_\mu$ and $(Z\alpha)^4m_\mu^2/M$ respectively) and at low $Z$ the Lamb shift is a dominant correction to the energy levels.

\section*{Hydrogen Lamb shift}

\begin{figure} 
\begin{minipage}[b]{0.45\textwidth}
\epsfig{file=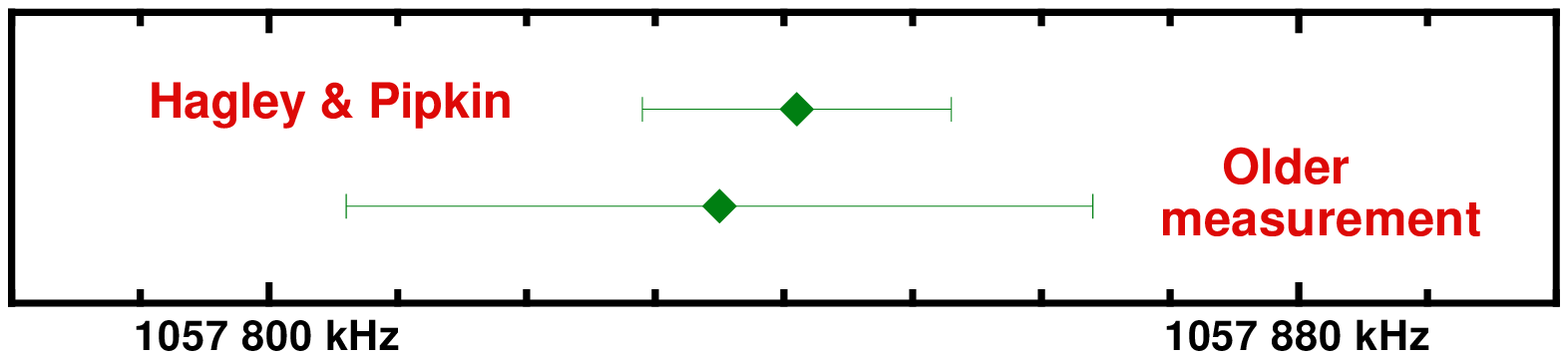,scale=0.41}
\end{minipage}%
\hskip 0.03\textwidth
\begin{minipage}[b]{0.45\textwidth}
\epsfig{file=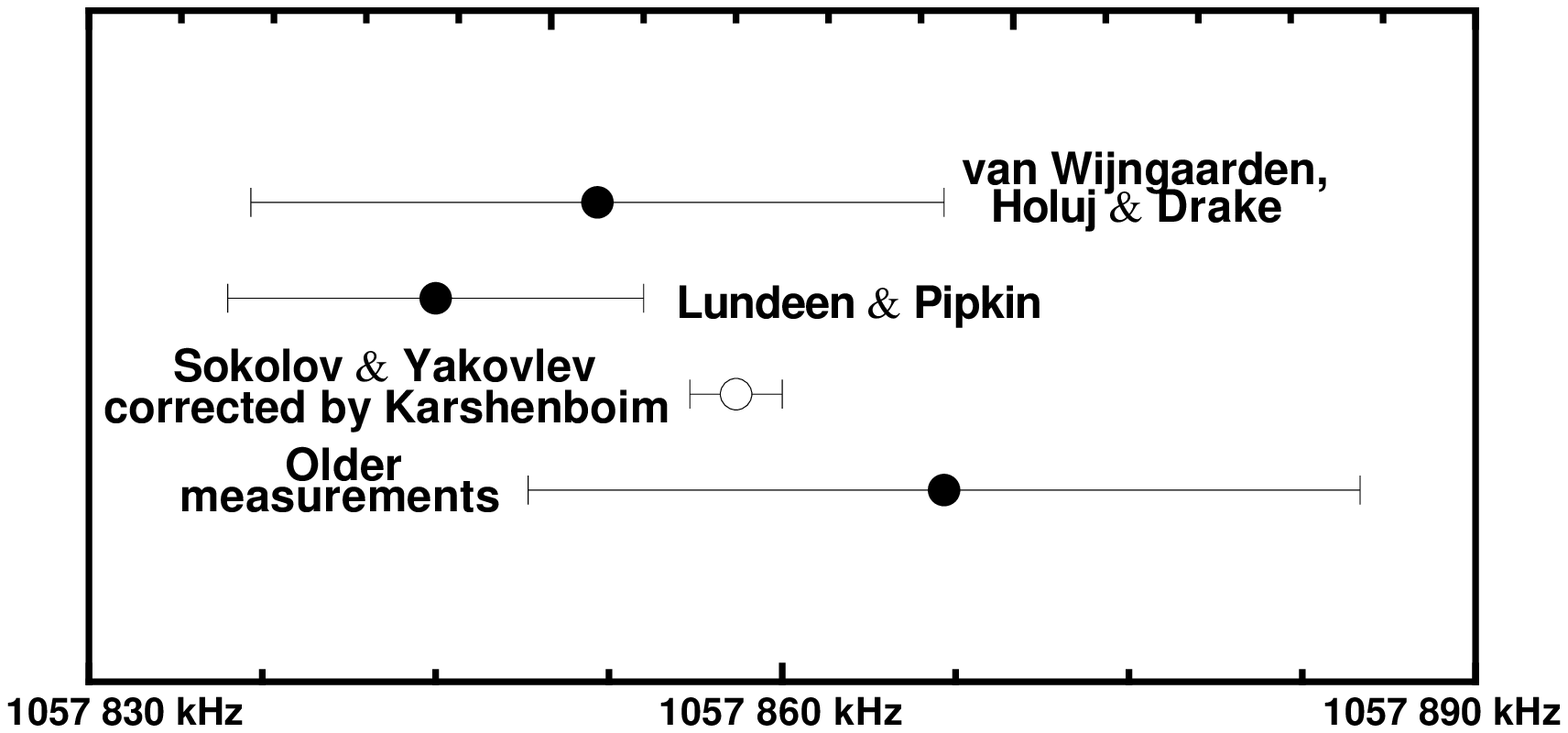,scale=0.41}
\end{minipage}
\vspace{10pt}
\begin{minipage}[t]{0.45\textwidth}
\caption{
Indirect determination of the Lamb shift (2s$_{1/2}$--2p$_{1/2}$) in atomic hydrogen via a study of the fine structure 2p$_{3/2}$--2s$_{1/2}$. See Ref. [3] for references. }
\label{f:sgk2}
\end{minipage}%
\hskip 0.08\textwidth%
\begin{minipage}[t]{0.45\textwidth}
\caption{
Direct measurements of the Lamb shift in the hydrogen atom. The references can be found in Ref. [3].}
\label{f:sgk3}
\end{minipage}
\end{figure}

A number of different splittings have been precisely studied for about a century. Bound state QED and maybe even QED itself was essentially established after a study of the Lamb shift and the fine and hyperfine structures in hydrogen, deuterium and helium ions. In the last decades, progress with such measurements was quite slow. The results of the last twenty years are presented in Fig 2 (Lamb shift) and 3 (fine structure recalculated in terms of the Lamb shift), while the older experiments are averaged (see Ref. \cite{CJP} for references). To reach the Lamb shift from the fine structure (2p$_{3/2}$--2s$_{1/2}$) measurement we need to use a value of the 2p$_{3/2}$--2p$_{1/2}$ splitting which was found theoretically. The most direct results of the Lamb shift need no QED theory. A result claimed to be the most accurate one ({\em Sokolov}$^\dag$) has an uncertainty of about 2 ppm. It is corrected because of a recalculation of the lifetime of the 2p$_{1/2}$ state \cite{2p}:
\[
\tau^{-1}(2p_{1/2})=\frac{2^{10}\pi}{3^8}\,\alpha^3\,
R_\infty\,\frac{m_R}{m}
\left\{1+\ln{\left(\frac{9}{8}\right)}\,\big(Z\alpha\big)^2
+\frac{\alpha\big(Z\alpha\big)^2}{\pi}\, \big(8.045...\big)\,
\ln{\frac{1}{\big(Z\alpha\big)^2}}
\right\}\,.
\]
There is some criticism by E. Hinds \cite{Hinds} and it is not clear if this result is as accurate as claimed. We wish to note, however, that common opinion on the direct Lamb shift measurement contains two contradicting statements. Firstly, it is generally believed that a Lamb splitting of 2s$_{1/2}$ and 2p$_{1/2}$ (about 1 GHz), with a decay width of 2p being 0.1 GHz, cannot be measured better than 10 ppm. This means that the statistic error should be larger than 10 ppm. Secondly, it is believed that Sokolov's experiment is incorrect only because of a possible systematic error claimed by Hinds. However, nobody insists that the statistical treatment of Sokolov's data was incorrect and we can hope that traditional methods can go far beyond 10 ppm level. Measurement of the deuterium Lamb shift within the Sokolov scheme will provide a chance to test some systematics of his experiment.

Essential progress in study of the hydrogen Lamb shift comes recently from the optical two--photon Doppler--free experiments (see {\em H\"ansch}$^\dag$ and {\em Schwob et al.}$^\dag$ fort detail). The Doppler--free measurement offers a determination of some transition frequency in the gross structure with a accuracy high enough to use the results to find the Lamb shift. However, two problems arise due to these experiments. 

\begin{figure}[ht] 
\begin{minipage}[b]{0.45\textwidth}
\centerline{\epsfig{file=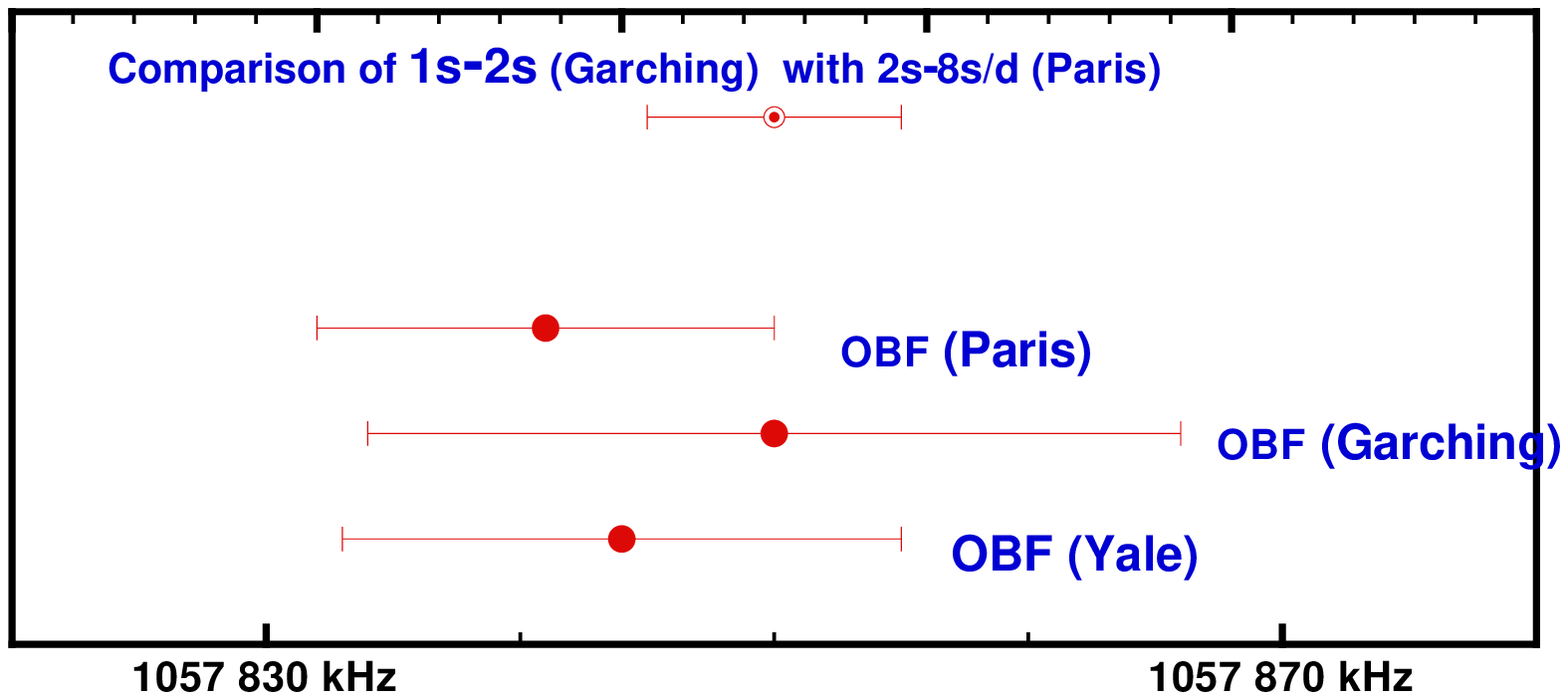,scale=0.45,bb=104 490 566 702}}
\end{minipage}%
\hskip 0.08\textwidth
\begin{minipage}[b]{0.45\textwidth}
\centerline{\epsfig{file=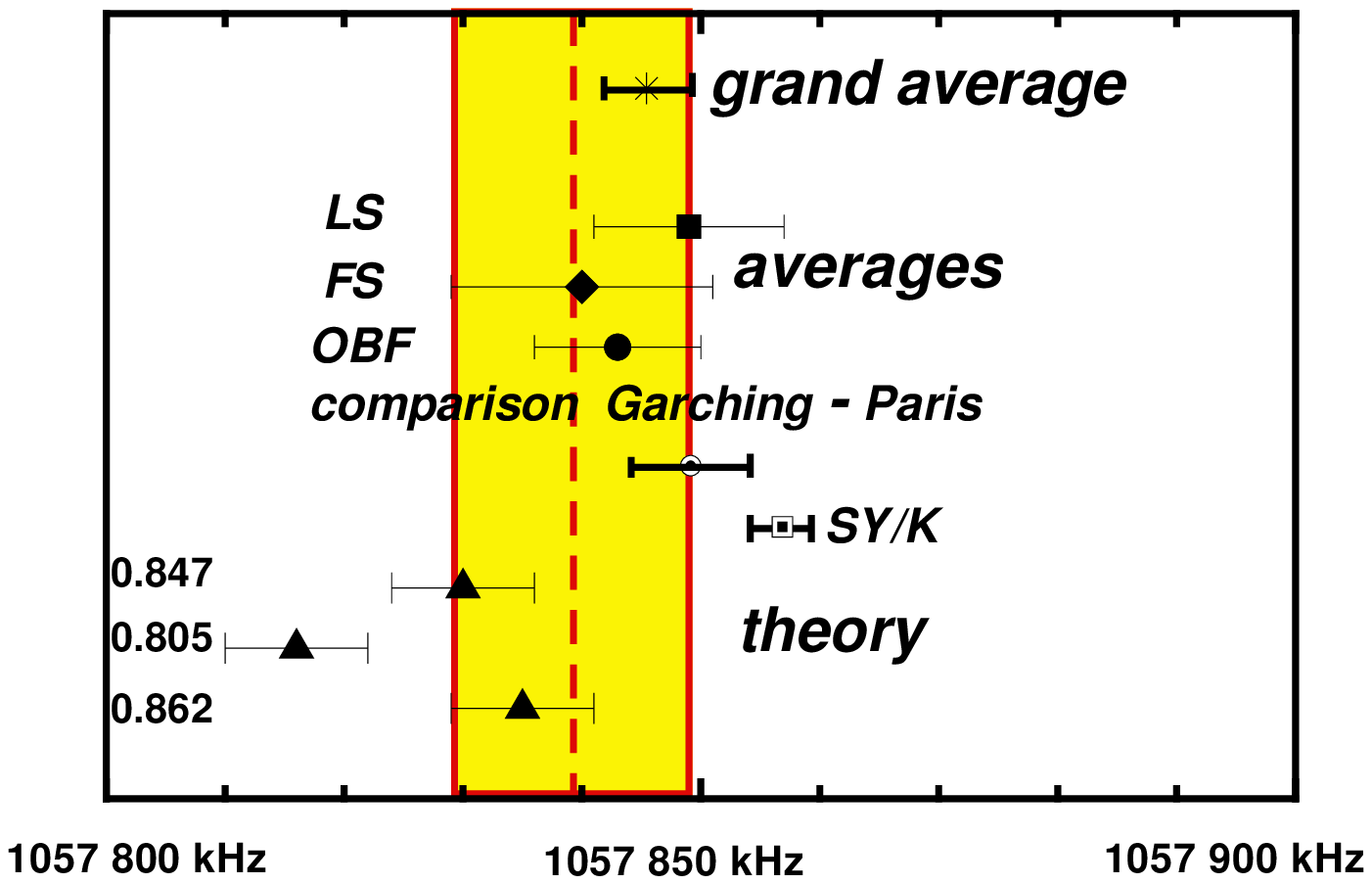,scale=0.45}}
\end{minipage}
\vspace{10pt}
\begin{minipage}[t]{0.45\textwidth}
\caption{
Optical determination of the lamb shift in the hydrogen atom. The references to the {\em optical beat frequency experiments} can be found in Ref. [3]. 
}
\label{f:sgk4}
\end{minipage}%
\hskip 0.08\textwidth
\begin{minipage}[t]{0.45\textwidth}
\caption{
Comparison of experiment and theory for the hydrogen Lamb shift. The references can be found in Ref. [3]. 
}
\label{f:sgk5}
\end{minipage}
\end{figure}

The transition energy between different levels of the gross structure is mainly determined by the Rydberg energy: $-R_\infty/n^2$. To extract the Lamb shift we first need to find a value of the Rydberg constant. There are two ways to manage this. Following the first of them one has to measure two different frequencies within one experiment with the ratio of the frequencies being an integer number. Obtaining a beat frequency one can avoid the problem of determining the Rydberg constant. Three experiments have been performed in this way: the Garching experiment dealt with the 1s--2s transition and the 2s--4s (and 4d), at Yale 1s--2s frequency was compared with the one--photon 2s-4p transition and that was the only precision optical experiment with a one--photon transition. The recent Paris experiment worked with 1s--3s and 2s--6s (and 6d). The values derived from these experiments are collected in Fig. 4.

Another way to manage the problem with the Rydberg constant is to do two independent absolute frequency measurements (i. e. measurements in respect to the primary cesium standard) and to compare them afterwards, hence determining both the Rydberg constant and the Lamb shift. Such an approach, combining results from Garching (1s--2s) and from Paris (on 2s--8s, --8d, --12d), gave another optical value (Fig. 4). Some of the optical experiments were also performed for deuterium and that may improve the accuracy in the determination of the Rydberg constant and, thence, of the hydrogen Lamb shift.

However, the values in Fig. 4 derived from the optical measurements need further theoretical treatment. The experiments involved a number of levels (1s, 2s, 3s etc) and with optical experimental data there was also a problem of an increasing number of levels with an unknown Lamb shifts. The problem was solved with the help of a specific difference \cite{JETP94}:
\begin{eqnarray}\label{Delta}
\Delta(n)&=&E_L(1s)-n^3\,E_L(ns)\nonumber\\
&=& \frac{\alpha(Z\alpha)^4}{\pi}\frac{m_R^3}{m^2}\times
\Biggl\{- \frac{4}{3}\ln{\frac{k_0(1s)}{k_0(ns)}}
\left(1+Z\frac{m}{M}\right)^2 +C_{Rec}\frac{Zm}{M}\nonumber\\
&+&(Z\alpha)^2
\Biggl[A_{61}\ln{\frac{1}{(Z\alpha)^2}}+
A^{VP}_{60}(n)+G^{SE}_{n}(Z\alpha) 
+\frac{\alpha}{\pi}\ln^2{\frac{1}{(Z\alpha)^2}}\,B_{62}(n) \Biggr]\Biggr\}\,,
\end{eqnarray}
where the coefficients $A_{61}$, $ A^{VP}_{60}(n)$, $ C_{Rec}$ and $ B_{62}(n)$ and a table for $ G^{SE}_{n}(Z\alpha)$ can be found in Refs. \cite{ZP97,CJP}. The uncertainty was also discussed there.
The difference has a better established status than that for 1s (or 2s) Lamb shifts (see Table 1). The uncertainty budget was improved recently after calculations of one--loop corrections, exactly at $Z=1$, ({\em Jentschura et al.}$^\dag$) and leading three--loop contributions ({\em Melnikov and van Ritbergen}$^\dag$). 

The theory of 2p$_{3/2}$--2p$_{1/2}$ splitting is also well established. Perhaps, we have to clarify here the word ``theoretical''. A value is a {\em theoretical} one if it is sensitive to {\em theoretical} problems (like the problem of the proton radius and of higher--order QED corrections for the Lamb shift). An insensitive, sterile value is not theoretical, it is rather a {\em mathematical} one, and that is the case for the difference $\Delta(2)$ and the 2p$_j$ energy. Details of theoretical calculations can be found in the review \cite{Eides}.

\begin{table}[ht]
\renewcommand{\arraystretch}{1.4}
\caption{Theoretical unceratinty of the different corrections for the Lamb shift in hydrogen. $^*$ In case of recoil term we present a value of contradiction between different calculations. $^\star$ We give an estimated unceratinty of proper reevaluation of the most accurate data.
}
\label{Ttheory}
\vskip 10pt
\begin{tabular}{cccc}
Contribution & $\delta E(2s)$ & $\delta \Delta(2)$ & $\delta E(2p)$ \\
\tableline
Two--loop  & 2 kHz & 0.6 kHz & 0.1 kHz\\
Recoil & 0.9 kHz$^*$ & - & -.\\
Radiative-recoil & 0.05 kHz & 0.05 kHz & 0.05 kHz \\
Nuclear structure &  $\sim$ 10 kHz$^\star$ & - & - \\
\end{tabular}
\end{table}

\section*{Nuclear structure effects}

Now we can compare theory and experiment for the 2s Lamb shift. We summarize them in Fig. 5, where we present average values for the Lamb shift, fine structure and optical beat frequency and comparison experiments. What is important is the influence of the nuclear charge distribution on the energy levels
\begin{equation}
\Delta E (nl) =
\frac{2}{3}\,\frac{(Z\alpha)^4}{n^3}\, m^3 \, R_N^2\;\delta_{l0}\;,
\end{equation}
where $R_N$ is a mean--squared nuclear charge radius. The position of theoretical values depends on the accepted value for the proton charge radius. We label three theoretical values with the proton radius (0.847 fm -- Mainz dispersion analysis paper, 0.805 fm -- Stanford scattering experiment, 0.862 fm -- Mainz scattering). More values for the proton radius are collected in Fig. 7 (see \cite{CJP} for references). To discuss the discrepancy let us look at the most important data on electron--proton elastic scattering presented in Fig. 8. One can see that the Mainz experiment is more appropriate to precisely determine the proton radius containing more points at lower momentum transfer and with a higher precision. Due to this any compilation containing the Mainz data has to lead to a result close to the Mainz result, because the Mainz scattering points must be statistically responsible for the final result and, in particular, the dispersion analysis performed by Mainz theorists led to such a result. However it ($R_p=0.847(9)$ fm) differs from the empirical value ($R_p=0.862(12)$ fm). One problem in evaluating the data is their normalization. One can write a low momentum expansion of the form factor
\begin{equation}
G(q^2)=a_0 + a_1 q^2 + a_2 q^4 + \dots
\end{equation}
From a theoretical point of view $G(0)=1$ indeed. However, the normalization measurement was accurate not enough (an in particular in the Mainz case it is about 0.5\%) and that means that a value tabulated from the data, as being the form factor, differed from it with some normalization. Three different fitting were performed by Wong \cite{Wong} (see Fig. 8). The free fittings of $a_0$ led to a larger uncertainty (Wong--Mainz value in Fig. 7). Even this result must be treated as a preliminary value. It is necessary to take into account some higher--order corrections and that is not possible because of the absence of any complete description of the experiment. The reasonable estimate of the theory is presented in Fig. 5 as a filled area. All experimental values are consistent with the theory exept the corrected value from the Sokolov and Yakovlev experiment. The present status is that the computation uncertainty is about 2 ppm, the measurement inaccuracy of the grand average value is 3 ppm, while the uncertainty due to the proton size is about 10 ppm.

\begin{figure}[ht] 
\begin{minipage}[b]{0.45\textwidth}
\centerline{\epsfig{file=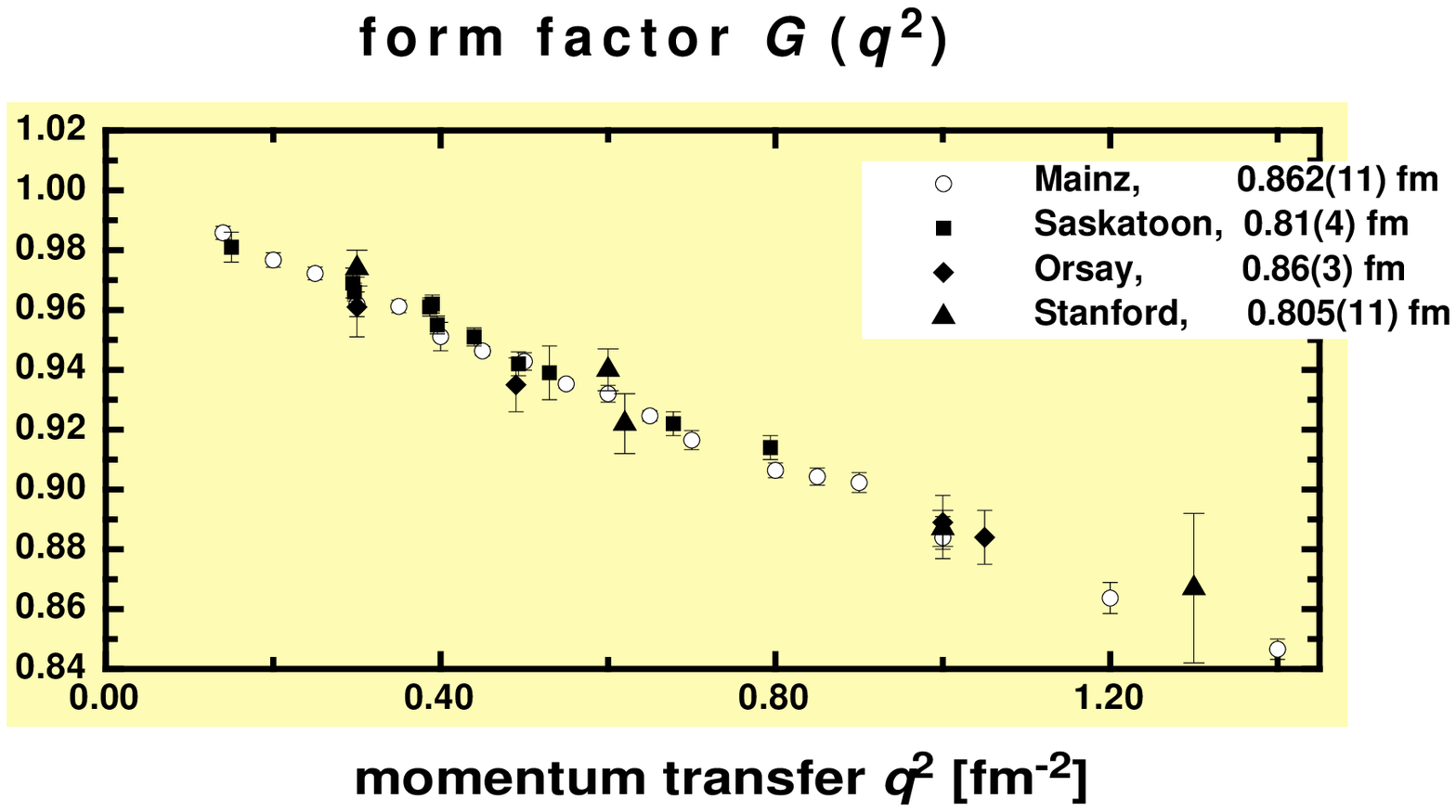,scale=0.45}}
\end{minipage}%
\hskip 0.08\textwidth
\begin{minipage}[b]{0.45\textwidth}
\centerline{\epsfig{file=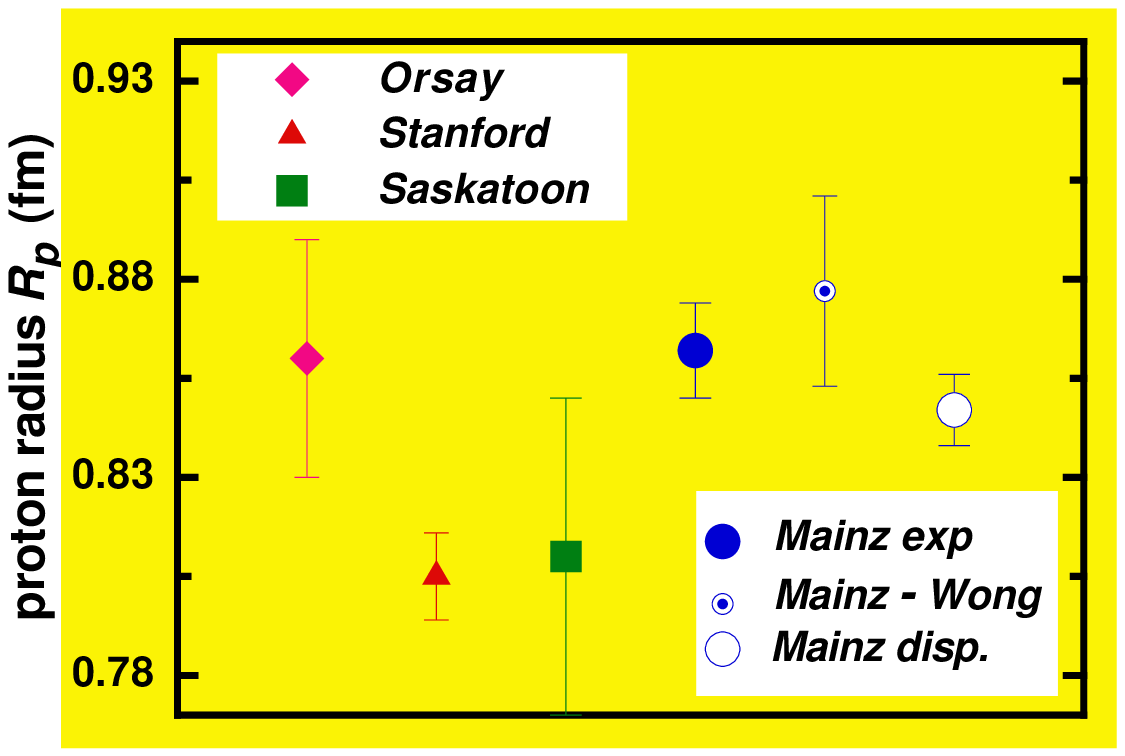,scale=0.45,bb=84 207 406 420}}
\end{minipage}
\vspace{10pt}
\begin{minipage}[t]{0.45\textwidth}
\caption{Electric form factor of the proton measured in
the scattering experiments (see Ref. [3] for references). 
}
\label{f:sgk6}
\end{minipage}%
\hskip 0.08\textwidth
\begin{minipage}[t]{0.45\textwidth}
\caption{
Proton charge radius determined from the scattering experiments. The references can be found in Ref. [3]. 
}
\label{f:sgk7}
\end{minipage}
\end{figure}

\begin{figure}[ht] 
\centerline{\epsfig{file=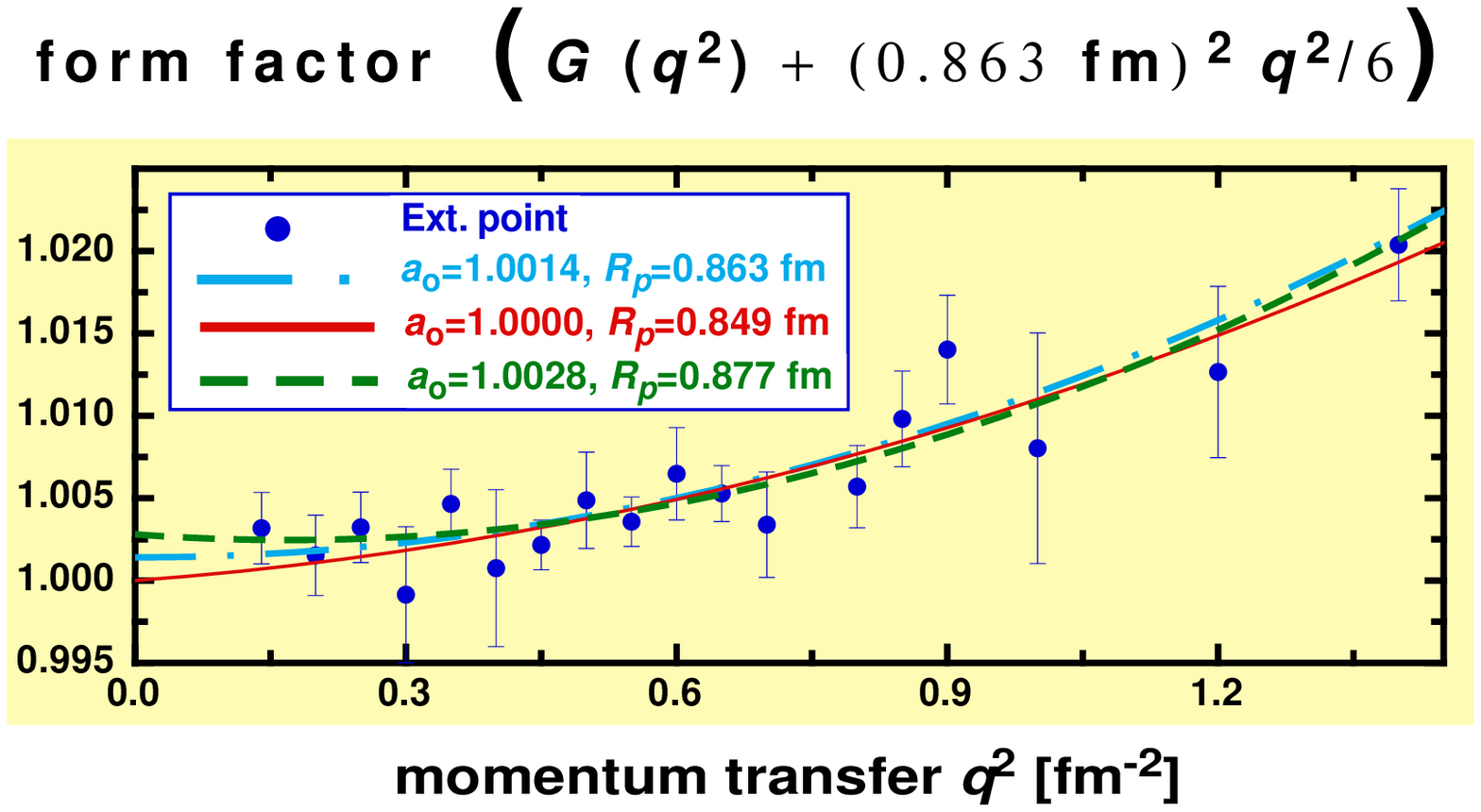,width=12cm}}
\vspace{10pt}
\caption{
Fitting the electric form factor of the proton from the Mainz experimental data. The references can be found in Ref. [3]. 
}
\label{f:sgk8}
\end{figure}

The problem of the nuclear size is not only a problem of the hydrogen Lamb shift: a similar situation arises with the helium--4 ion Lamb shift, where uncertainties resulting from the QED computation and the nuclear size are about the same. The comparison of theory to experiment is presented in Fig. 9. The evolution of the measured value has been due to a study of possible systematic sources ({\em Drake and van Wijngaarden}$^\dag$).

\begin{figure}[ht] 
\begin{minipage}[b]{0.29\textwidth}
\centerline{\epsfig{file=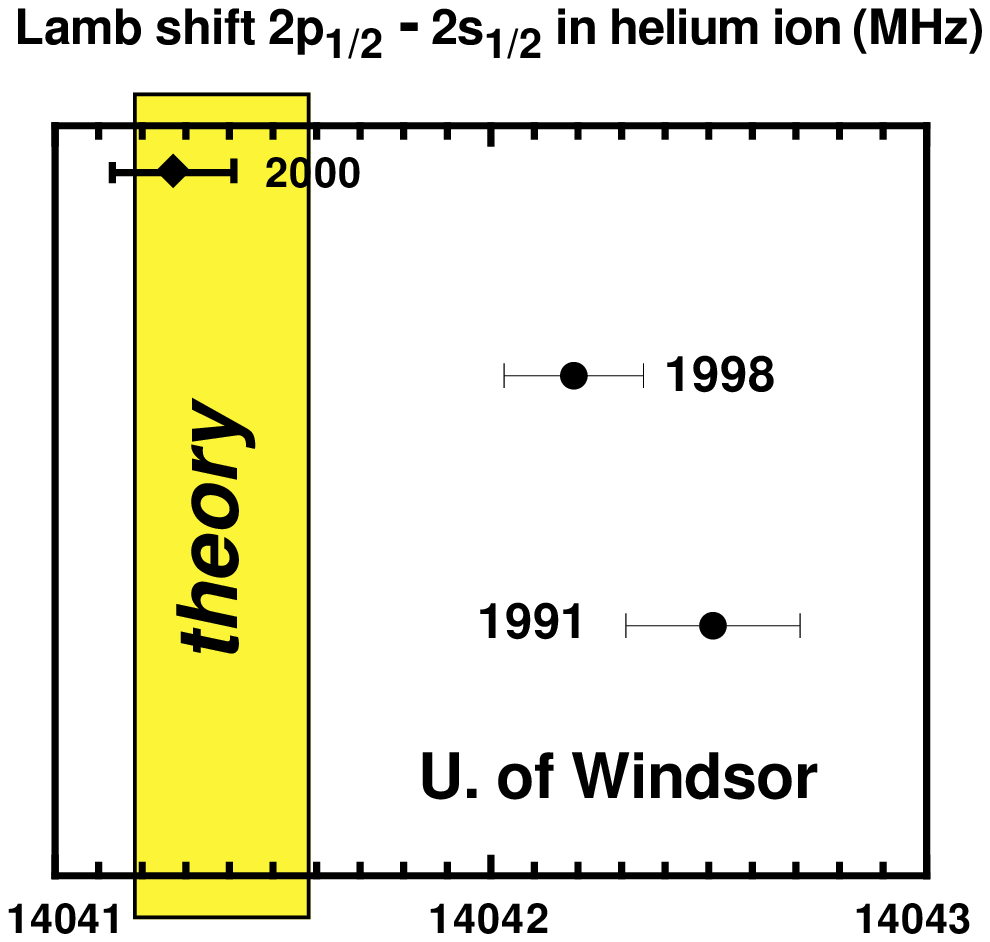,scale=0.4}}
\end{minipage}%
\hskip 0.05\textwidth
\begin{minipage}[b]{0.29\textwidth}
\centerline{\epsfig{file=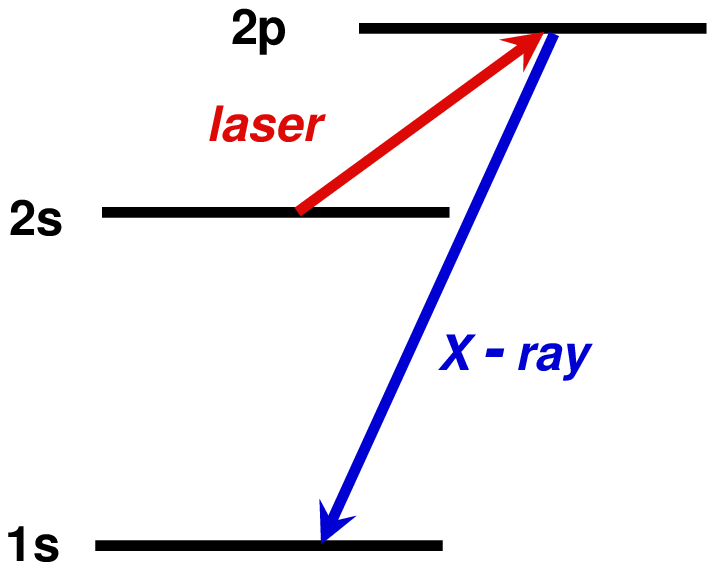,scale=0.4,bb=125 --17 328 150}}
\end{minipage}%
\hskip 0.05\textwidth
\begin{minipage}[b]{0.29\textwidth}
\centerline{\epsfig{file=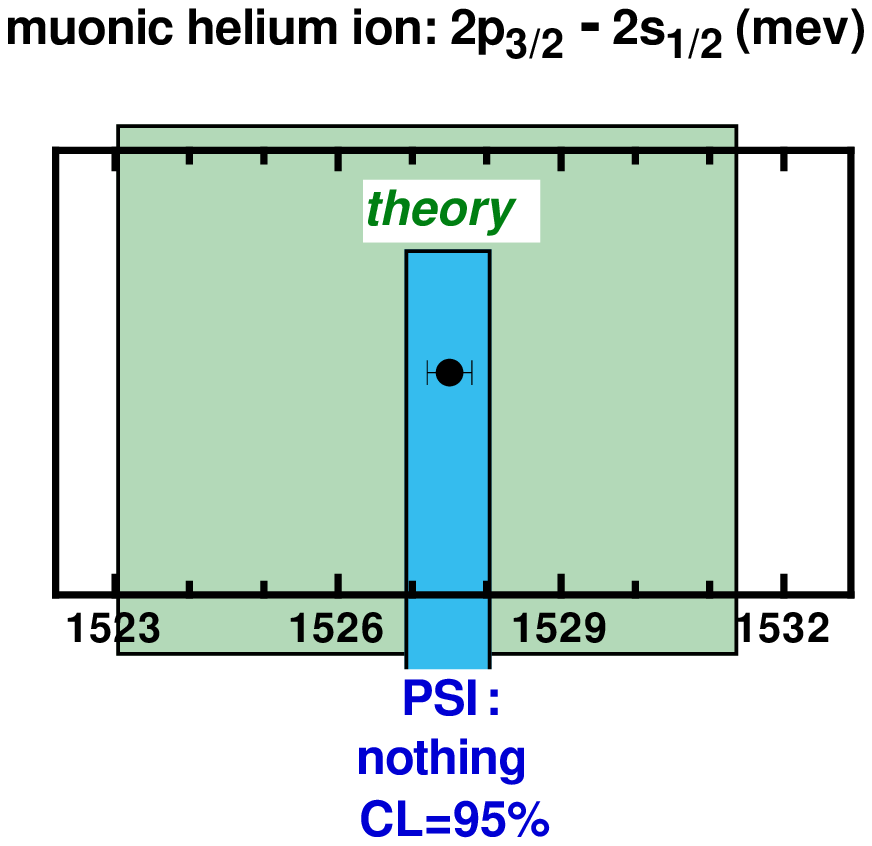,scale=0.4}}
\end{minipage}
\vspace{10pt}
\begin{minipage}[t]{0.29\textwidth}
\caption{
Experiment on the Lamb shift in ions of
helium--4 at the University of Windsor ({\em Drake and Wijngaarden$^\dag$}).
}
\label{f:sgk9}
\end{minipage}%
\hskip 0.05\textwidth
\begin{minipage}[t]{0.29\textwidth}
\caption{
Scheme of the experiment on the Lamb shift
in a light muonic atom
}
\label{f:sgk10}
\end{minipage}%
\hskip 0.05\textwidth
\begin{minipage}[t]{0.29\textwidth}
\caption{Study of the Lamb shift in muonic helium.}
\label{f:sgk11}
\end{minipage}
\end{figure}

\section*{Proton--free hydrogen physics}

A more difficult problem is that of the hyperfine structure, which is more sensitive to the nuclear structure. While the experimental uncertainty is below $10^{-12}$, the theoretical inaccuracy is about 10 ppm. The main problem is a distribution of the magnetic moment inside the proton. It seems the scattering cannot provide accurate enough data and we need to discuss how to manage the problem of nuclear structure by means of the atomic physics. We consider here three ways to do that:
\begin{itemize}
\item one is based on study of muonic atoms (the muon is a lepton with a lifetime of about 2 $\mu$s and a mass of about $207 \,m_e$);
\item another deals with the special difference $\Delta_{hfs}=E_{hfs}(1s)-E_{hfs}(2s)$ (cf. in Eq. (\ref{Delta})), which can be precisily measured;
\item the third is for atoms without nuclear structure. In such an atom one must substitute the proton by some more appropriate positive particle (muon or positron).
\end{itemize}
A promising way is to determine the nuclear structure with muonic atoms and in particular with muonic hydrogen. The muon orbit lies lower than the electron one. Since $m_\mu\simeq 207\, m_e$ the muon hydrogen Bohr radius is about 200 times smaller than that in hydrogen and, hence, the former is more sensitive to nuclear effects. A scheme of an experiment running now at PSI ({\em Pohl et al}$^\dag$) is presented in Fig. 10. The experiment consists of the following steps: creating a metastable 2s state, exiting it to the 2p state by a laser, measuring the intensity of the X--ray decay 2p--1s.
A similar scheme was used for muonic helium \cite{muhelium}, however a recent experiment by PSI \cite{PSIhelium} showed no appropriate signal (Fig. 11). A study of the helium experiment revealed a crucial point: the creation of a metastable state, which can be destroyed by collisions. The collision rate is proportional to the target gas density as well as the rate of creating the muonic atom, and so the density cannot be varied arbitrarily. The slow muon beam at PSI allow one to use a low density gas target and creation of the 2s state has been detected. In case of success, the PSI experiment will give us the charge radius of the proton and the so--called Zemach correction to the 2s state of muonic hydrogen. Comparison of the muonic hydrogen hfs and hydrogen hfs will allow us to go farther with the study of the proton structure. 

Another way to manage the problem of nuclear structure is to compare the 1s and 2s hfs. The experiments were performed for hydrogen (recently by {\em Rothery and Hessels}$^\dag$), deuterium and helium ion. The recent hydrogen experiment has attracted our attention to the problem of $\Delta_{hfs}$ and it was discovered ({\em Karshenboim and Sapirstein}$^\dag$) that the results (and primarily those for the helium ion) are quite sensitive to higher order corrections. All value for the hfs used to be presented in units of Fermi energy ($\nu_F$), which is the result of a nonrelativistic interaction of the magnetic moment of an electron in the 1s state and the nucleus. The accuracy of the difference allows one to detect the fourth order corrections, namely, $\alpha(Z\alpha)^3$, $\alpha^2(Z\alpha)^2$, $\alpha(Z\alpha)^2m/M$, and $(Z\alpha)^3m/M$.

The same fourth order corrections are now a subject of study in the muonium ground state hfs (see Table 2). A muonium atom is a kind of hydrogen without the proton: instead the proton the nucleus is a positive muon. The present status of the muonium hyperfine splitting is as follows: the experiment at LAMPF gave 4 463 302 765(53) Hz , while the theoretical prediction is consistent with the experiment but less accurate. A computational part of the uncertainty is about 200 Hz, while a hidden experimental uncertainty in the theory is about 500 Hz. It is due to a calculation of the Fermi energy, which is proportional to the muon magnetic moment, determined from the same LAMPF experiment. Possible progress is considered by {\em Jungmann}$^\dag$.

\begin{table}[ht]
\renewcommand{\arraystretch}{1.4}
\caption{Fourth order corrections to the muonium hyperfine structure. Most of references can be found in Refs. [11] and [1].}
\label{Tfourth}
\vskip 10pt
\begin{tabular}{ccc}
Contribution & Numerical result & Reference \\
\tableline
$(Z\alpha)^4$   & 0.03 kHz   & Breit \\
$(Z\alpha)^2(m/M)^2\ln(1/(Z\alpha))$ & -0.11 kHz  & Lepage \\
&& Bodwin {\em et al}.\\
$\alpha^2(Z\alpha)(m/M)\ln^3(M/m)$ & -0.05 kHz  & Eides and Shelyuto \\
$\alpha^2(Z\alpha)(m/M)\ln^2(M/m)$ & 0.01 kHz  & Eides {\em et al}. \\
$\alpha(Z\alpha)^2(m/M)\ln^2(1/(Z\alpha))$ & 0.34 kHz  & Karshenboim \\
$(Z\alpha)^3(m/M)\ln^2(1/(Z\alpha))$ & -0.04 kHz  & Karshenboim \\
$(Z\alpha)^3(m/M)\ln(M/m)\ln(1/(Z\alpha))$ & -0.21 kHz  & Karshenboim \\
&& Kinoshita and Nio\\
$\alpha^2(Z\alpha)^2\ln^2(1/(Z\alpha))$ & -0.04 kHz  & Karshenboim \\
$\alpha(Z\alpha)^3\ln(1/(Z\alpha))$ & -0.47 kHz  & Karshenboim \\
$\alpha(Z\alpha)^{3+}$ (SE) & -0.04 kHz  & Blundell {\em et al}.\\
$\alpha(Z\alpha)^3$ (VP) & 0.02 kHz  & Karshenboim {\em et al}.\\
$\alpha(Z\alpha)(m/M)^2$      &  -0.04    & Eides {\em et al}.\\
$(Z\alpha)^2(m/M)^2$      &  0.01    & Pachucki \\
\end{tabular}
\end{table}

\begin{figure}[ht] 
\begin{minipage}[b]{0.45\textwidth}
\centerline{\epsfig{file=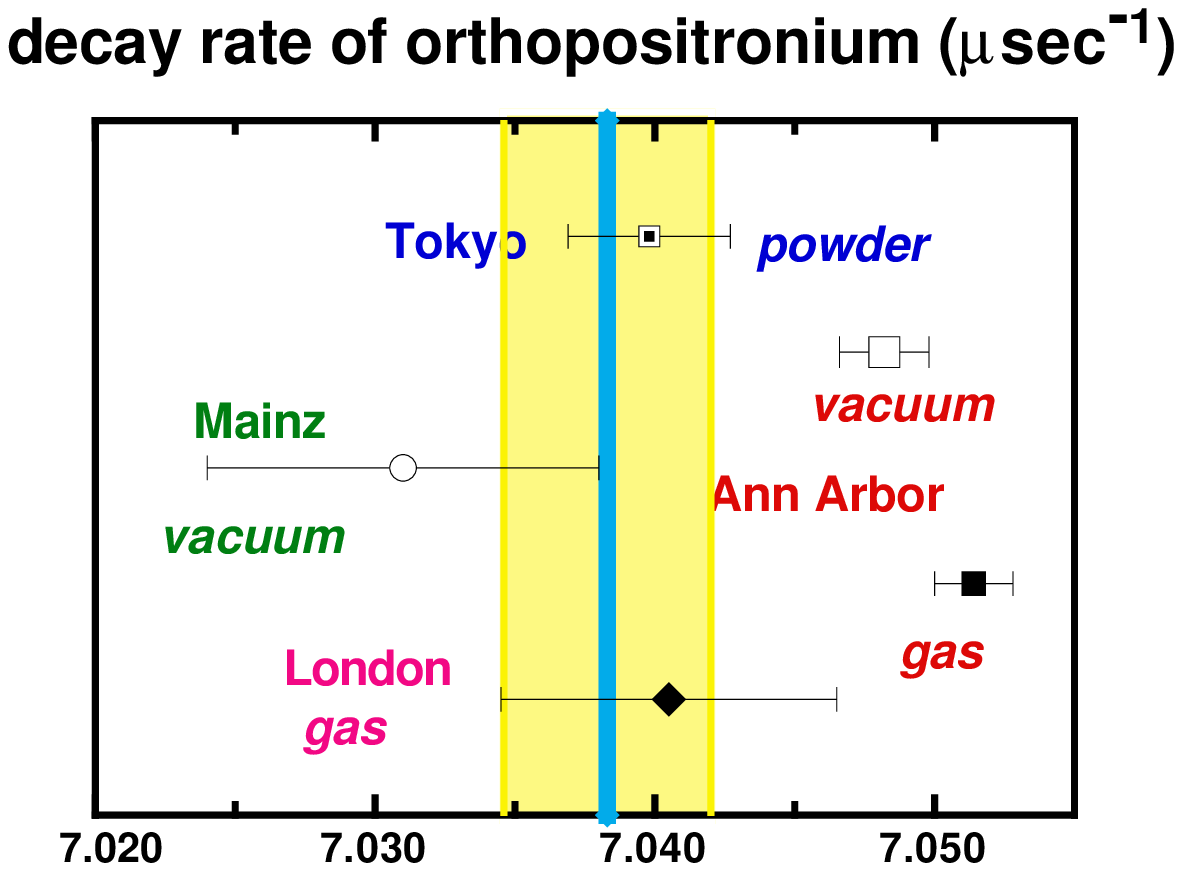,scale=0.45}}
\end{minipage}%
\hskip 0.08\textwidth
\begin{minipage}[b]{0.45\textwidth}
\centerline{\epsfig{file=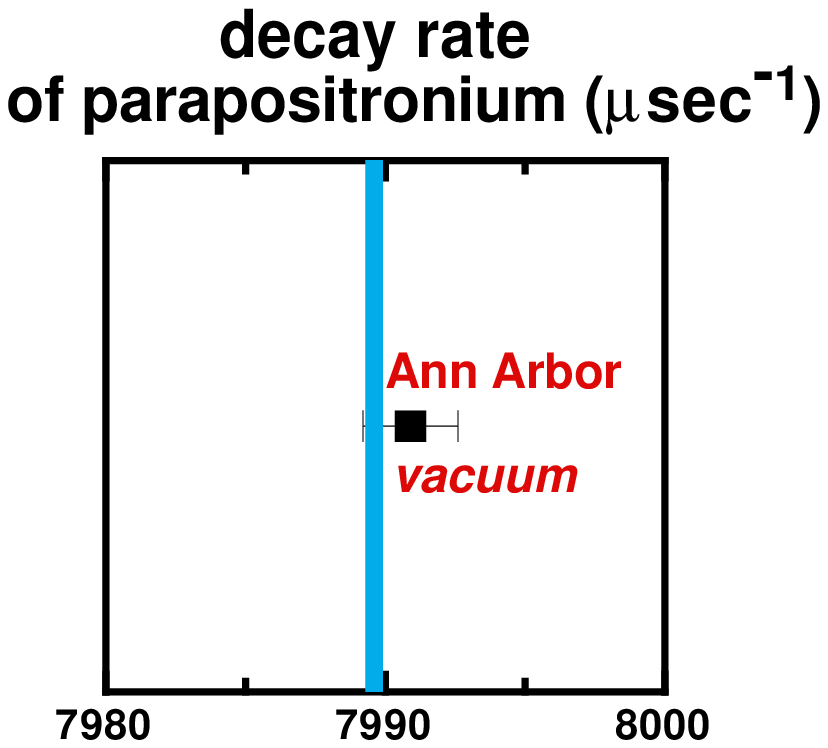,scale=0.45,bb=412 59 657 279}}
\end{minipage}
\vspace{10pt}
\begin{minipage}[t]{0.45\textwidth}
\caption{Decay of orthopositronium. The references can be found in {\em Conti}$^\dag$.}
\label{f:sgk12}
\end{minipage}%
\hskip 0.08\textwidth
\begin{minipage}[t]{0.45\textwidth}
\caption{Decay of parapositronium (see {\em Comti}$^\dag$ for reference).}
\label{f:sgk13}
\end{minipage}
\end{figure}

\begin{figure}[ht] 
\begin{minipage}[b]{0.45\textwidth}
\centerline{\epsfig{file=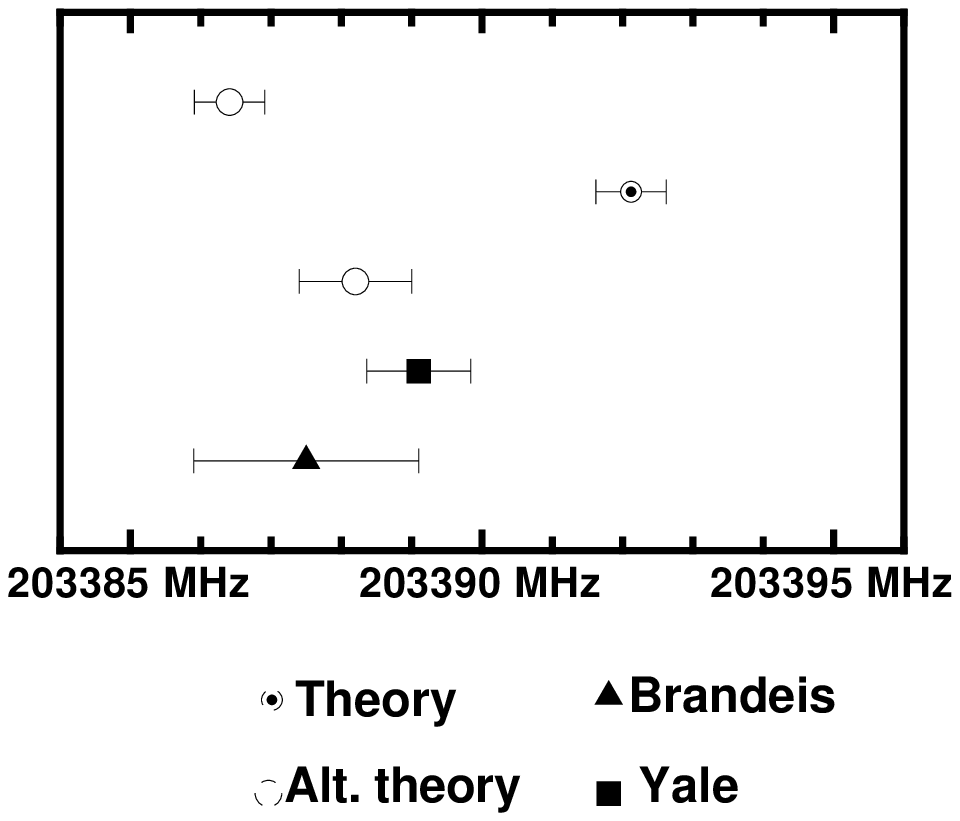,scale=0.41}}
\end{minipage}%
\hskip 0.08\textwidth
\begin{minipage}[b]{0.45\textwidth}
\centerline{\epsfig{file=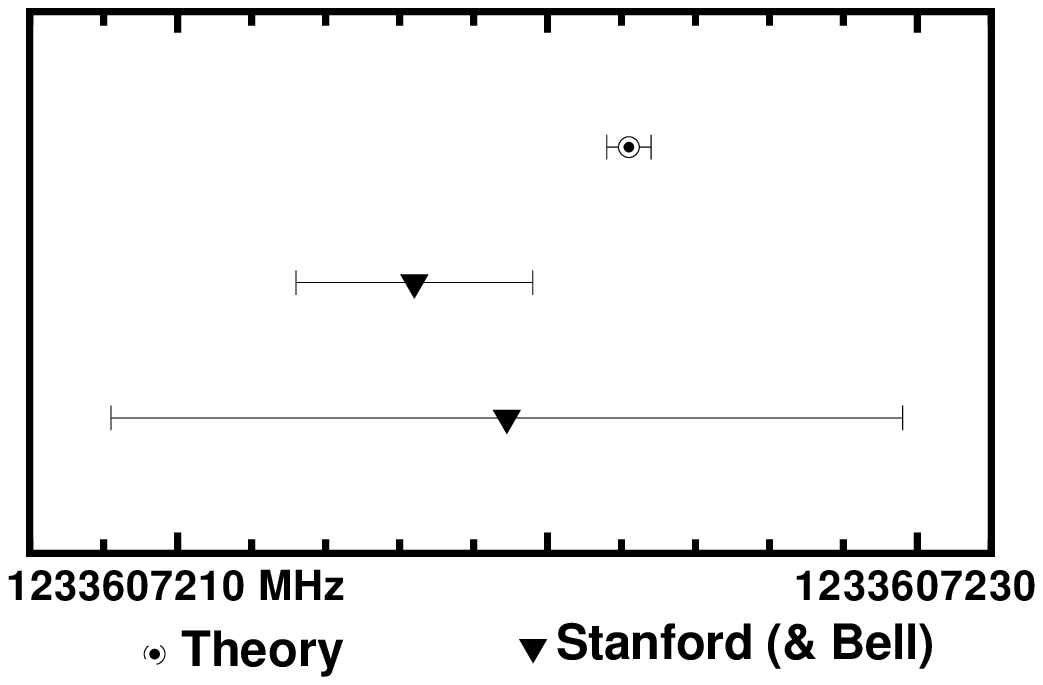,scale=0.41,bb=283 71 588 274}}
\end{minipage}
\vspace{10pt}
\begin{minipage}[t]{0.45\textwidth}
\caption{
Hyperfine structure interval in the ground state of the positronium atom. The references can be found in {\em Conti}$^\dag$.
}
\label{f:sgk14}
\end{minipage}%
\hskip 0.08\textwidth
\begin{minipage}[t]{0.45\textwidth}
\caption{
1s--2s transition in positronium: comparison of experiment with theory. The references can be found in {\em Conti}$^\dag$.
}
\label{f:sgk15}
\end{minipage}
\end{figure}
\begin{figure}[ht] 
\centerline{\epsfig{file=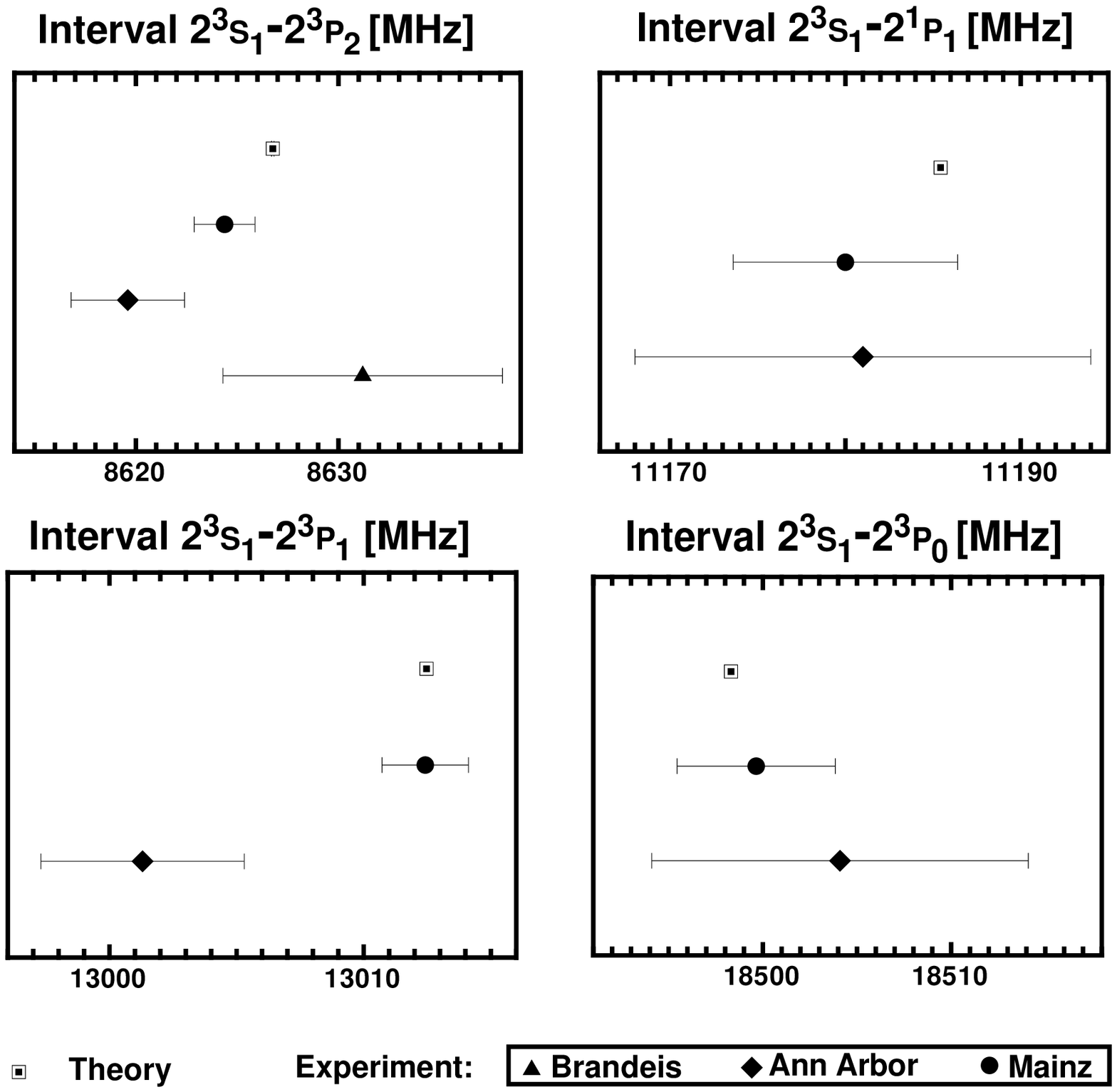,width=12cm}}
\vspace{10pt}
\caption{Positronium fine structure: theory and experiment. See  {\em Conti}$^\dag$ for references.}
\label{f:sgk16}
\end{figure}

Another proton--free simple atom is positronium. Its lifetime is much shorter than that of muonium, but it can be more easily produced. Different measurements in positronium are summarized in Fig. 12--16. Energy levels in positronium can be presented in the form
\[
R_\infty\times\left\{C_{20}+ C_{40}\alpha^2+ C_{50}\alpha^2+ 
\left(C_{61}\ln(1/\alpha)+ C_{60}\right)\alpha^2+\left(C_{72}\ln^2(1/\alpha)+ \dots\right)\alpha^3
\right\}\,.
\]
After two decades of intensive theoretical study we know the coefficients in the above expression up to $C_{60}$ (\cite{Adkins}, {\em Czarnecki et al.}$^\dag$) and $C_{72}$ (\cite{JETP93,Pachucki}). The decay width in positronium is known up to fractinal order $\alpha^2$ ({\em Czarnecki et al.}$^\dag$ for parapositronium) (and for orthopositronium {\em Adkins}$^\dag$) and $\alpha^3\ln^2\alpha$ \cite{JETP93}. Since the Adkins's result is a preliminary one we do not include it in Fig. 12. The Adkins's result led to a fractional correction about $4\alpha^2$ and so the theory is in contradiction to the Ann Arbor experiment. Some progress in the study of positronium is expected in the near future ({\em Conti}$^\dag$).

\section*{The status of bound state QED}

After briefly reviewing the studies for hydrogen, muonium and positronium, let us discuss a problem and current trend of bound state QED. First of all we need to mention that, in our mind, the QED theory as a theory is well established. As a pure theory it is absolutely correct and absolutely useless: 
\begin{itemize}
\item
The QED theory is a theory of interaction between leptons (electrons and muons) and photons only. We need to include hadrons (such as proton) into consideration. Even for the case of pure leptonic values (like for muonium) we need to calculate a hadronic vacuum polarization contribution. So the QED theory is incomplete. 
\item
The QED theory cannot predict anything exactly, but only in terms of expansion and the uncertainty can be presented in terms like $O(\alpha^7m)$. It is necessary to develop an effective approach to estimate uncalculated corrections quantitively in Hz and eV.
\item
The QED deals rather with free particles and it is necessary to develop an effective approach to solve a bound state problem for two bodies.
\end{itemize}
These three problems lie beyond QED as a mathematical theory, but are an essential part of any real QED calculations. A test of different effective approaches is a real problem, as is evaluating of hadronic contributions needed for the precision theory of simple atoms.

The bound state problem has mainly three small parameters: $\alpha$ (associated with QED effects), $Z\alpha$ (due to binding effects) and $m/M$ (the recoil parameter). Now, for the first time, it is necessary to try to really study effects which involve essential QED, two--body and binding effects simultaneously (the $\alpha(Z\alpha)^2m/M$ corrections). That is a problem for the hyperfine structure of muonium, the 1s/2s hyperfine structure in hydrogen, deuterium and He$^+$ and for the positronium spectrum. Another crucial problem is that of the Lamb shift in hydrogen and light ions: this is a higher order two--loop corrections ($\alpha^2(Z \alpha)^6m$) already known in part. We summarize all crucial terms in Table 3. 

\begin{table}[ht]
\renewcommand{\arraystretch}{1.4}
\caption{Crucial higher--order corrections in current studies of the simple atoms. For the $g$- factor of an electron we recalculated the corrections in terms of corrections to the energy.}
\label{Tcrucial}
\vskip 10pt
\begin{tabular}{cc}
Value & Order \\
\tableline
hydrogen gross structure         & $\alpha(Z\alpha)^7m$, $\alpha^2(Z\alpha)^6m$ \\
hydrogen fine structure              & $\alpha(Z\alpha)^7m$, $\alpha^2(Z\alpha)^6m$ \\
hydrogen Lamb shift              & $\alpha(Z\alpha)^7m$, $\alpha^2(Z\alpha)^6m$ \\
He$^+$ Lamb shift              & $\alpha(Z\alpha)^7m$, $\alpha^2(Z\alpha)^6m$ \\
nitrogen fine structure              & $\alpha(Z\alpha)^7m$, $\alpha^2(Z\alpha)^6m$ \\
$^3$He$^+$ hyperfine structure   & $\alpha(Z\alpha)^7m^2/M$, $\alpha^2(Z\alpha)^6m^2/M$, $\alpha(Z\alpha)^6m^3/M^2$, $(Z\alpha)^7m^3/M^2$\\
muonium hyperfine structure      & $\alpha(Z\alpha)^7m^2/M$, $\alpha(Z\alpha)^6m^3/M^2$, $(Z\alpha)^7m^3/M^2$\\
positronium hyperfine structure  & $\alpha^7m$ \\
positronium gross structure      & $\alpha^7m$ \\
positronium fine structure       & $\alpha^7m$ \\
parapositronium decay rate       & $\alpha^7m$ \\
orthopositronium decay rate      & $\alpha^8m$ \\
parapositronium $4\gamma$ decay  & $\alpha^8m$ \\
orthopositronium $5\gamma$ decay & $\alpha^8m$ \\
$g$--factor of electron in $^{40}$Ca$^{19+}$ & $\alpha(Z\alpha)^7m$, $\alpha^2(Z\alpha)^6m$ \\
$g$--factor of free electron             & $\alpha^8m$ \\
\end{tabular}
\end{table}

The three parameters we mention generate different expansions and it is found that all three kind of expansions are not free from problems.
\begin{itemize}
\item
 The QED expansion over $\alpha$ is an asymptotic one and the value of terms will decrease to some $n_c$ and increase after it. Fortunately, that is not important for $n=1-3$, which are only actual for the bound state QED.
\item
 The $Z \alpha$ expansion involves another problem. One knows that high $Z$ is a bad limit (strong coupling) and it is believed that low $Z$ is a good limit. The latter is wrong. It is clear that at $Z=0$ there is no bound system at all and the behavior of any expansion in the limit of low $Z$ is not an analytical one. This eventually leads to logarithmic contributions. Even a cube of logarithm ($\ln^3(1/\alpha) \sim 120$ at $Z=1$) appears. The imaginary part of the logarithm is $\pi$ and the non--leading terms have often large coefficients because of this. There is another mechanism for large coefficients. As it is well--known the Bethe logarithm ($\ln\big(k_0(ns)\big)\sim 3$) is a logarithm of an effective energy (in atomic units) of an intermediate state in a calculation of the electron self--energy. A logarithm equal to 3 corresponds to a quite relativistic intermediate p--state ($v/c \sim 4.5 (Z \alpha)$) and that also leads to large coefficients because of relativistic corrections for intermediate states.
\item
 The recoil effect with the $m/M$ expansion also involves a non--analytical behavior. It is correct that the limit $m_1=m_2$ (positronium) is rather complicated for a calculation, however at the opposite limit ($m/M\to 0$) there is no bound state. Hopefully, often the logarithmic recoil corrections are not quite important numerically. Since most of the recoil effects are relativistic ones, the exchange loop generates the effective parameter $(Z \alpha)/\pi$ rather than $Z \alpha$.
\end{itemize}
Due to the increasing number of logarithmic contributions we end up with a problem of large higher--order corrections. Some higher--order logarithmic terms are compatible in comparison to a constant part of some lower--order terms. 
\begin{itemize}
\item For the hydrogen 2s Lamb shift, non--logarithmic parts of the fifth order corrections (in unit of the Rydberg contributions) lie from 164 kHz ($\alpha(Z\alpha)^6m$) and 37 kHz ($\alpha^2(Z\alpha)^6m$) to a few kHz for recoil terms. The leading logarithmic term in the next order is $\alpha^2(Z\alpha)^6\ln^3(Z\alpha)$, which contributs 3.6 kHz.
\item The non--logarithmic parts of the third order correction (in unit of $\nu_F$) for the ground state muonium hyperfine splitting varies from 8.8 kHz ($\alpha(Z\alpha)^2$) to 2 kHz for recoil and radiative recoil terms and to 0.4 kHz for $\alpha^2(Z\alpha)$. Three leading logarithmic corrections are slightly below 1 kHz (see Table 2): $\alpha(Z\alpha)^3$, $\alpha(Z\alpha)^2(m/M)$ and $(Z\alpha)^3(m/M)$.
\item A number of positronium levels are under study. The non-logarithmic $\alpha^6m$ term (7.2 MHz) for the hyperfine structure is bigger than the logarithmic part of the $\alpha^7m$ contribution (0.9 MHz), while for the 1s-2s interval the situation is different: the non-logarithmic $\alpha^6m$ term is only 0.5 MHz and that is less than 1.2 MHZ of the $\alpha^7m$.
\end{itemize}
These example show that an estimation of the higher order terms is extremely important and we hope that a calculation of leading logarithmic contributions provides a reasonable way to estimate uncalculated terms. We estimate the non--leading term within a half--value of the leading logarithmic contributions. 

Estimation of uncalculated terms is a crucial problem in any QED calculations. Let us now mention the case of moderate Z. Study of these ions provides a unique possibility of measuring higher order corrections. In particular an experimental study of helium ({\em Drake}$^\dag$, {\em Burrows et al.}$^\dag$) and nitrogen ({\em Myers}$^\dag$) hydrogen--like ions will allow us to extract information on higher--order two--loop contributions with the help of a theoretical study of all other terms ({\em Ivanov and Karshenboim}$^\dag$). Moderate--$Z$ few--electron atoms allow us to test our understanding of higher--order electron--electron interactions which is important for high $Z$ spectroscopy. 

Large values of the higher--order terms imply a calculation without expansion. This is only possible for one parameter, either $Z \alpha$ or $m/M$. For the simplest corrections (like e. g. the vacuum polarization) it is possible to calculate analytically to any order of $Z\alpha$, otherwise only numerical results are possible. It is unlikely that a complete exact calculation of the two-loop self-energy can be performed soon and that means that expansion techniques is still the main approach in calculating the higher-order terms, perhaps in  combination with experiments. 

A new opprortunity appears due to recent measurements of a bound electron $g$--factor in a hydrogen--like atom ({\em H\"affner et al.}$^\dag$, \cite{Quint}). A recent result on the carbon ion is useful for indirect determination of the electron mass ({\em Karshenboim}$^\dag$). Our theoretical prediction
\begin{equation}
 g_b(e) = 2\cdot\big(1+520795(1)\cdot10^{-9}\big)
\end{equation}
mainly based on \cite{Beier} calculation has a smaller uncertainty in part because of taking into account a known $\alpha^2(Z\alpha)^2$ term. Studies of the $g$--factor will be very different from the Lamb shift and the hyperfine structure. In contrast to spectroscopic studies it is possible to go through all $Z$ and to determine some unknown coefficients of the theoretical expansion if we can fix its shape (we call this {\em weak theory} in contrast to a real theory which can give direct numerical predictions).

\section*{Summary}

Concluding the paper we wish to mention briefly different applications for the study of simple atoms. These studies are important for different field of physics:
\begin{itemize}
\item Determination of fundamental constants ($R_\infty$, $\alpha$, $m_e/m_p$): some of these are important for other application, like e. g. the fine structure constant is necessary to reproduce the value of the Ohm from the quantum Hall effect.
\item Development of new optical standard and tool for same: like e. g. the new frequency chain designed recently (see {\em Diddam et al.}$^\dag$, {\em Udem et al.}$^\dag$ and \cite{Holzwarth}). 
\item {\em New} physics: a study of muonium-to-antimuonium conversion ({\em Jungmann}$^\dag$) and the exotic decay of positronium ({\em Conti}$^\dag$) provide a possibility of looking for new particles, while the antihydrogen project and some others are expected to test some symmerties or possible variations of fundamental constants.
\item Particle physics: a recent study of hydrogen is rather important to learn the proton structure, than to test the bound state QED. The theoretical methods developped recently are of use both for atoms and two-quark particles (mesons) and QED is a good opprotunity to test the methods. Exotic atoms give us accurate information on hadron-hadron interactions.
\item Nuclear physics: the situation is similar to that for particles physics: a study of light atoms offers information on structure of their nuclei. One the other hand, two-- and three--body atoms are an appropriate problem to testing different effective methods before to applying them to light nuclei.
\end{itemize}

Most of these questions and a more broad range of problems in physics of
simple atoms were considered at a Satellite meeting to the ICAP (Hydrogen
Atom, 2: {\sl Precise Physics of Simple Atomic System}). The Proceeding will be published by Springer in 2001.

\section*{Aknowledgement}

I am grateful to T. H\"ansch, K. Jungmann, G. Werth, J. Sapirstein and E. Myers for useful and stimulating discussions. I would like to thank S. Nic Chormaic and F. Cataliotti for useful remarks on manuscript. The work was supported in part by RFBR (grant 00-02-16718), NATO (CRG 960003) and Russian State Program {\em Fundamental Metrology}.

\end{document}